\documentclass[aps,prb,twocolumn,10pt,superscriptaddress,amsfonts,amssymb,amsmath,preprintnumbers,floatfix,noeprint]{revtex4-2}
\usepackage{graphicx}
\usepackage{bm}
\usepackage{color,ulem}
\usepackage[usenames,dvipsnames]{xcolor}
\usepackage[dvipdfmx,colorlinks=true,citecolor=ProcessBlue,urlcolor=blue,linkcolor=ProcessBlue]{hyperref}

\begin{document}
\title{Observation of spin-conserving two-spinon continuum in the $S$=1/2 antiferromagnetic chain system Sr$_2$CuO$_3$ using Cu $K$-edge resonant inelastic x-ray scattering}
\author{Kenji Ishii}
\affiliation{Synchrotron Radiation Research Center, National Institutes for Quantum Science and Technology, Hyogo 679-5148, Japan}
\author{Kenji Tsutsui}
\affiliation{Synchrotron Radiation Research Center, National Institutes for Quantum Science and Technology, Hyogo 679-5148, Japan}
\author{Takayuki Kawamata}
\thanks{Present address: Department of Natural Sciences, Tokyo Denki University, Tokyo 120-8551, Japan}
\affiliation{Department of Applied Physics, Tohoku University, Sendai 980-8579, Japan}
\author{Yoji Koike}
\affiliation{Department of Applied Physics, Tohoku University, Sendai 980-8579, Japan}
\date{\today}

\begin{abstract}
We report a Cu $K$-edge resonant inelastic x-ray scattering (RIXS) study of spin excitations in the $S$=1/2 antiferromagnetic chain system Sr$_2$CuO$_3$.
The spectral weight observed below the charge-transfer gap appears in two-spinon continuum, indicating the fractionalization of a spin-conserving ($\Delta S = 0$) magnetic excitation into two-spinon states.
The intensity of these excitations reaches a maximum near the midpoint between the center and the boundary of the Brillouin zone, and decreases toward the zone boundary;  this behavior contrasts with that of the spin-flip ($\Delta S = 1$) excitations typically observed via inelastic neutron scattering or Cu $L_3$-edge RIXS.
The momentum dependence of the intensity is described by the spin-exchange dynamical structure factor.
A phenomenological analysis of the symmetry between the polarization and the $d$ orbital explains the resonance condition for the two-spinon excitations.
\end{abstract}

\preprint{preprint \today}

\maketitle

\section{Introduction}

Low-dimensional $S$=1/2 antiferromagnetic Heisenberg systems, which are often realized in complex copper oxides, have attracted considerable attention in condensed matter physics.
Two-dimensional systems serve as a playground for high-temperature superconductivity and a wide variety of materials have been synthesized.
On the other hand, in the one-dimensional $S$=1/2 chain, strong quantum fluctuations prevent the long-range antiferromagnetic order, and a quantum spin liquid state emerges.
The physical properties of the one-dimensional system are studied in the materials, such as Sr$_2$CuO$_3$ and SrCuO$_2$.
In these real materials, long-range order occurs due to weak interchain interactions; the Neel temperature of Sr$_2$CuO$_3$ is $\sim$5 K \cite{Keren1993,Kojima1997}.
Even so, the characteristic properties of the one-dimensional system remain.

\begin{figure*}[t]
\centering
\includegraphics[width=1.0\textwidth]{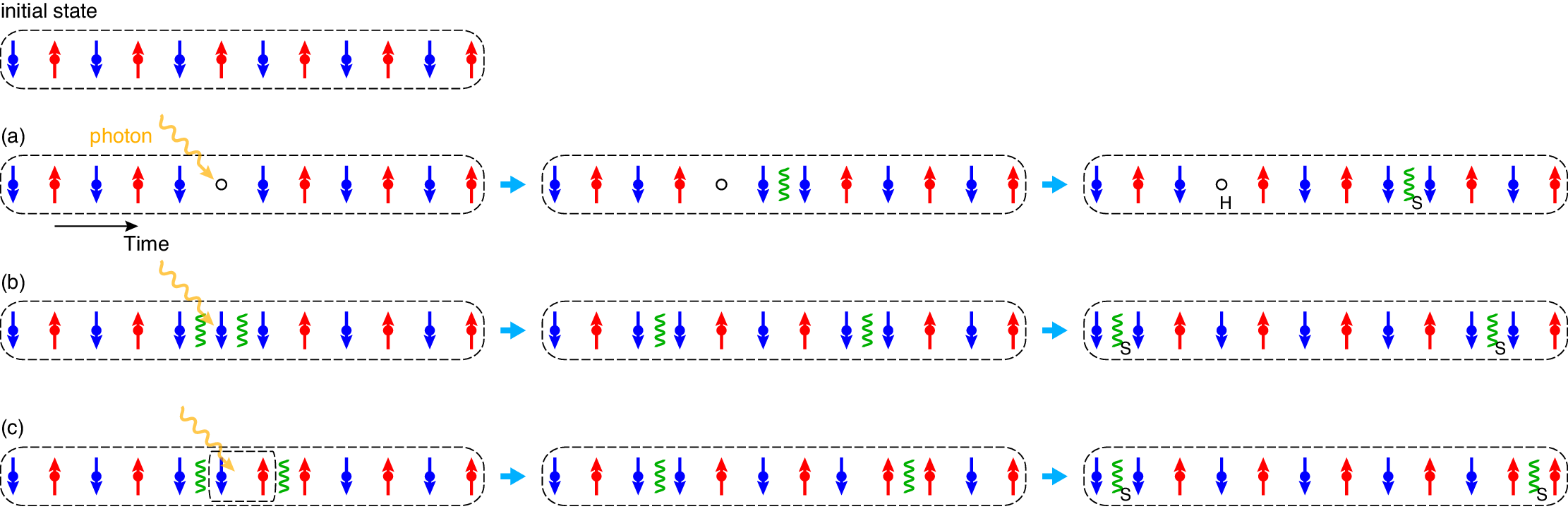}
\caption{Fractionalization of elementary excitations in the one-dimensional spin chain.
(a) Removal of an electron by photon irradiation breaks into a holon (H) and a spinon (S).
(b) A single spin-flip excitation ($\Delta S = 1$) splits into two spinons.
(c) A double spin-flip and total spin-conserving excitation ($\Delta S = 0$) is also fractionalized into two spinons.
The wavy vertical line represents an antiferromagnetic domain boundary where the nearest-neighbor spins are aligned in parallel.}
\label{fig:fractionalization}
\end{figure*}

One of the remarkable properties of the one-dimensional spin chain is the fractionalization of the elementary excitations.
The removal of an electron from the $S$=1/2 antiferromagnetic chain breaks into a holon and a spinon, and they propagate independently, as shown in Fig.~\ref{fig:fractionalization}(a).
This phenomenon called spin-charge separation was observed using angle-resolved photoemission spectroscopy \cite{Kim1996,Kim1997,Neudert1998,Fujisawa1999,Kim2006}.
When a single spin-flip ($\Delta S = 1$) excitation is created, another type of fractionalization emerges as two spinons [Fig.~\ref{fig:fractionalization}(b)], forming a two-spinon continuum in the spin excitation spectrum.
Recently, the fractionalization of an electron transition from one orbital to another into a spinon and an orbiton was found as a spin-orbital separation \cite{Wohlfeld2011,Schlappa2012,Wohlfeld2013,Bisogni2015}.
The spinons in the one-dimensional copper oxides are responsible for the intriguing transport properties, such as ballistic thermal conduction \cite{Takahashi2006,Kawamata2008} and spin current \cite{Hirobe2017}.

The two-spinon continuum was originally measured using the inelastic neutron scattering (INS) \cite{Stone2003,Zaliznyak2004,Lake2005,Kohno2007,Walters2009,Lake2010}, 
where scattering intensity is determined by the single-spin dynamical structure factor.
Since the 2010s, resonant inelastic x-ray scattering (RIXS) using brilliant synchrotron radiation x-rays has emerged as an alternative technique for observing the spinon continuum \cite{Schlappa2012,Bisogni2015,Fumagalli2020,Li2021}.
Similar to the INS, the intensity of the cross-polarized Cu $L_3$-edge RIXS is also proportional to the single-spin dynamical structure factor of materials with Cu$^{2+}$ under a fast-collision approximation where the core-hole lifetime is very short \cite{Ament2009}.

Historically, the observation of new types of magnetic excitations has marked important milestones in the development of RIXS.
Two-magnon excitations in the parent compounds of copper oxide superconductors were the first magnetic excitations observed using RIXS at both the O $K$-edge \cite{Harada2002} and the Cu $K$-edge \cite{Hill2008}.
Later on, it was demonstrated that RIXS at the $L$-edge has sensitivity to single spin-flip excitations in the two-dimensional antiferromagnetic copper oxide La$_2$CuO$_4$, wherein spin-flip excitations emerge as dispersive magnons in the energy-momentum space \cite{Braicovich2010}.
Almost at the same time, single $J_{\rm eff}=1/2$ pseudospin-flip excitations in an iridium oxide have also  been reported \cite{JKim2012}.
These observations of the single spin-flip excitations are major breakthroughs of RIXS.
Thanks to the significant improvement in energy resolution, the RIXS technique has been employed to investigate momentum-resolved magnetic excitations in a variety of materials \cite{Zhou2013,Fabbris2017,Brookes2020,Lu2021a,Suzuki2021,Suzuki2023}.
After the capability to observe the spin-flip excitations was demonstrated, RIXS was applied to the one-dimensional spin systems, where the RIXS intensity agrees with the two-spinon excitations \cite{Glawion2011,Schlappa2012,Bisogni2014,Li2021}.
The RIXS study of the one-dimensional system was further extended to higher-order multi-spinons \cite{Schlappa2018,Kumar2022}.

As mentioned above, $\Delta S = 1$ excitations in the one-dimensional $S$=1/2 antiferromagnetic chain form a two-spinon continuum in the excitation spectrum, which is described by the single-spin dynamical structure factor.
The upper ($E_{\rm u}$) and lower ($E_{\rm l}$) boundaries of the continuum are given respectively by
\begin{eqnarray*}
E_{\rm u} &=& \pi J \sin (\pi q) \\
E_{\rm l} &=& \frac{\pi J}{2} \sin (2\pi q),
\end{eqnarray*}
where $J$ is the exchange interaction between the nearest-neighbor spins, and $q$ denotes the momentum transfer along the chain in reciprocal lattice unit \cite{Caux2006}.
It is noted that a substantial four-spinon contribution is present within the two-spinon continuum of the single-spin dynamical structure factor \cite{Caux2006,Klauser2011a,Mourigal2013}.

Interestingly, it has been theoretically predicted that a double spin-flip and total spin-conserving ($\Delta S = 0$) excitation is also fractionalized into two spinons \cite{Klauser2011a,Forte2011a}, and that this fractionalization can be probed by RIXS at the $K$-edge.
A spinon can be regarded as a magnetic domain boundary, and the difference between single and double spin-flip excitations lies in the relative spin directions at the boundary.
The direction of the spins at the two boundaries is parallel for the $\Delta S = 1$ excitation [Fig.~\ref{fig:fractionalization}(b)], while it is antiparallel for the $\Delta S = 0$ excitation [Fig.~\ref{fig:fractionalization}(c)].
The excitation of two spinons induced by the $\Delta S = 0$ process is described by the spin-exchange dynamical structure factor.
This dynamical structure factor represents the leading term of RIXS for magnetic excitations with $\Delta S = 0$ under the ultrashort core-hole lifetime (UCL) approximation.
The continuum of the spin-exchange dynamical structure factor spans the same energy-momentum space as that of the single-spin dynamical structure factor, but the two exhibit different intensity distributions.
At first glance, one might expect that each spin flip fractionalizes into two spinons, leading to a total of four spinons for a double spin-flip ($\Delta S = 0$) excitation.
However, these theoretical studies have demonstrated that nearly all $\Delta S = 0$ excitations in the spin-exchange dynamical structure factor fractionalize into two spinons, in contrast to the $\Delta S = 1$ excitations in the single-spin dynamical structure factor.
Calculated RIXS spectra beyond the UCL approximation, obtained using the Hubbard and Heisenberg models, are consistent with the spin-exchange dynamical structure factor \cite{Kourtis2012,Igarashi2012a}, even though a direct comparison has not yet been performed.

In this paper, we apply RIXS at the Cu $K$-edge to examine the two-spinon continuum induced by the $\Delta S = 0$ excitations experimentally.
The advantage of the Cu $K$-edge over the Cu $L_3$-edge is a selection rule of the spin excitations; the Cu $K$-edge RIXS is insensitive to the $\Delta S = 1$ excitations, while the $\Delta S = 1$ excitations overlap with $\Delta S = 0$ excitations in the Cu $L_3$-edge RIXS spectra.
It is noted that polarization analysis of the scattered x-rays is effective in separating the $\Delta S = 0$ excitations from the $\Delta S = 1$ excitations \cite{Braicovich2014,Minola2015}, although the use of the polarization analysis is still limited.
The O $K$-edge RIXS is also insensitive to the $\Delta S = 1$ excitations but it has limitations in the accessible momentum space.
Taking these advantages of the Cu $K$-edge, we observe RIXS spectral weight of the two-spinon continuum and the momentum dependence of the spectral weight agrees with the spin-exchange dynamical structure factor.
Additionally, we calculate the spin-exchange dynamical structure factor and the Cu $K$-edge RIXS spectra using the same cluster model, and confirm that the distribution of the spectral weight is consistent between them.

\section{Experimental Details}

A single crystal of Sr$_2$CuO$_3$ was grown using the traveling-solvent floating-zone technique.
The crystal was cut so that each surface was perpendicular to the crystallographic $a$, $b$, and $c$-axes, and x-rays were irradiated on one of the surfaces.
The Cu $K$-edge RIXS experiments were conducted at beam line 11XU of SPring-8.
The incident photons were monochromatized using a Si(111) double crystal monochromator, and the energy width was further reduced using either an asymmetric four-bounce Si(333) or Si(400) monochromator.
A spherically-bent diced Ge(733) analyzer was employed to analyze the scattered photon energy.
The total energy resolutions using the Si(333) and Si(400) monochromators were 55--63 and 95 meV, respectively.
The sample was kept at the base temperature of a cryostat ($\sim$10 K) throughout the measurements.

\section{Results}

\subsection{Incident photon energy dependence}

\begin{figure}[t]
\centering
\includegraphics[scale=1.0]{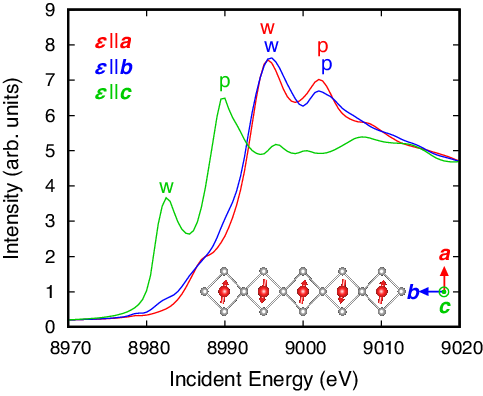}
\caption{Polarization-dependent x-ray absorption spectra of Sr$_2$CuO$_3$.
Each spectrum has two peaks near the absorption edge and they correspond to the well-screened (w) and poorly-screened (p) core-hole final states.
The inset shows a one-dimensional array of Cu-O plaquettes along with the crystallographic axes.}
\label{fig:abs}
\end{figure}

\begin{figure}[t]
\centering
\includegraphics[scale=0.8]{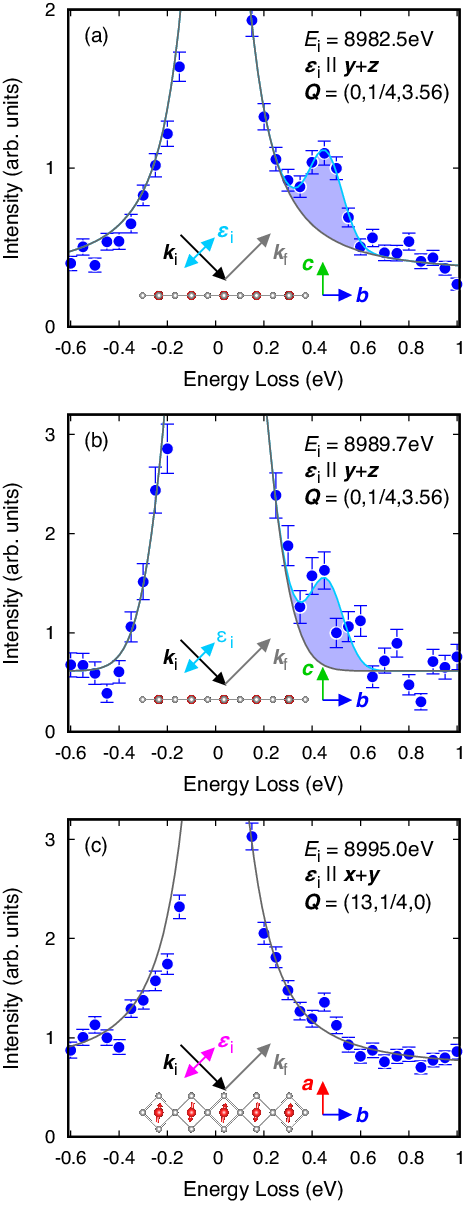}
\caption{Cu $K$-edge RIXS spectra of Sr$_2$CuO$_3$ taken at $q = 1/4$.
The incident photon energy ($E_\mathrm{i}$), incident photon polarization (${\boldsymbol \epsilon}_{\rm i}$), absolute momentum transfer $\bm{Q}$ are shown in each figure, where ${\bm x}$, ${\bm y}$ and ${\bm z}$ represent the unit vectors along the crystallographic ${\bm a}$, ${\bm b}$ and ${\bm c}$ directions, respectively.
The scattering geometry and crystal orientation are shown in the inset, in which $\bm{k}_\mathrm{i}$ and $\bm{k}_\mathrm{f}$ denote the incident and scattered photon wavevector, respectively.}
\label{fig:pol}
\end{figure}

Figure \ref{fig:abs} shows the x-ray absorption spectra measured in total fluorescence mode.
The photon polarization (${\boldsymbol \epsilon}$) is parallel to one of the crystallographic axes shown in the inset.
These spectra are characteristic of Cu atoms in a planar four-fold coordination and closely resemble those of T'-structured Pr$_2$CuO$_4$ \cite{Oyanagi1990} and Nd$_2$CuO$_4$ \cite{Tan1990}.
The spectrum appears at lower energies when the polarization is perpendicular to the Cu–O plaquettes (${\boldsymbol \epsilon} \parallel {\bm c}$).
Each spectrum exhibits two peaks near the absorption edge.
The lower-energy peak corresponds to the well-screened core-hole final state.
This state has a significant $\vert \underline{1s}4p3d^{10}\underline{L} \rangle$ character, in which electron transfer from the O $2p$ orbital to the Cu $3d$ orbital effectively screens the $1s$ core hole.
Here, $\underline{1s}$ and $\underline{L}$ denote a Cu $1s$ core hole and an O $2p$ hole, respectively.
The higher-energy peak arises from a transition to the poorly screened core-hole final state, which predominantly has $\vert \underline{1s}4p3d^{9} \rangle$ character \cite{Kosugi1989}.
We select 8982.5 eV (well-screened states of ${\boldsymbol \epsilon} \parallel {\bm c}$), 8989.7 eV (poorly-screened states of ${\boldsymbol \epsilon} \parallel {\bm c}$), and 8995.0 eV (well-screened states of ${\boldsymbol \epsilon} \parallel {\bm a}, {\bm b}$) as incident photon energies ($E_\mathrm{i}$) in the following RIXS measurements.

Figure \ref{fig:pol} shows RIXS spectra measured at the different incident photon energies using the Si(333) monochromator.
Momentum transfer along the chain ($b$-axis) is kept at 1/4 while that perpendicular to the chain is chosen so that the scattering angle ($2\theta$) is close to 90$^{\circ}$ to minimize the elastic scattering.
Here, absolute momentum transfer ($\bm{Q}$) is expressed in reciprocal lattice units with the lattice constants $a$ = 12.6910 \AA, $b$ = 3.9089 \AA, and $c$ = 3.4940 \AA~\cite{Ami1995}, where the Cu-O plaquettes are stacked along the $c$-axis.
Incident photon polarization (${\boldsymbol \epsilon}_\mathrm{i}$) and crystal orientation are shown in the inset.
Because of the choice of $E_\mathrm{i}$, only the ${\boldsymbol \epsilon}_\mathrm{i} \parallel {\bm c}$ component contributes to the resonance in the configurations (a) and (b).

\begin{figure*}[t]
\centering
\includegraphics[width=0.8\textwidth]{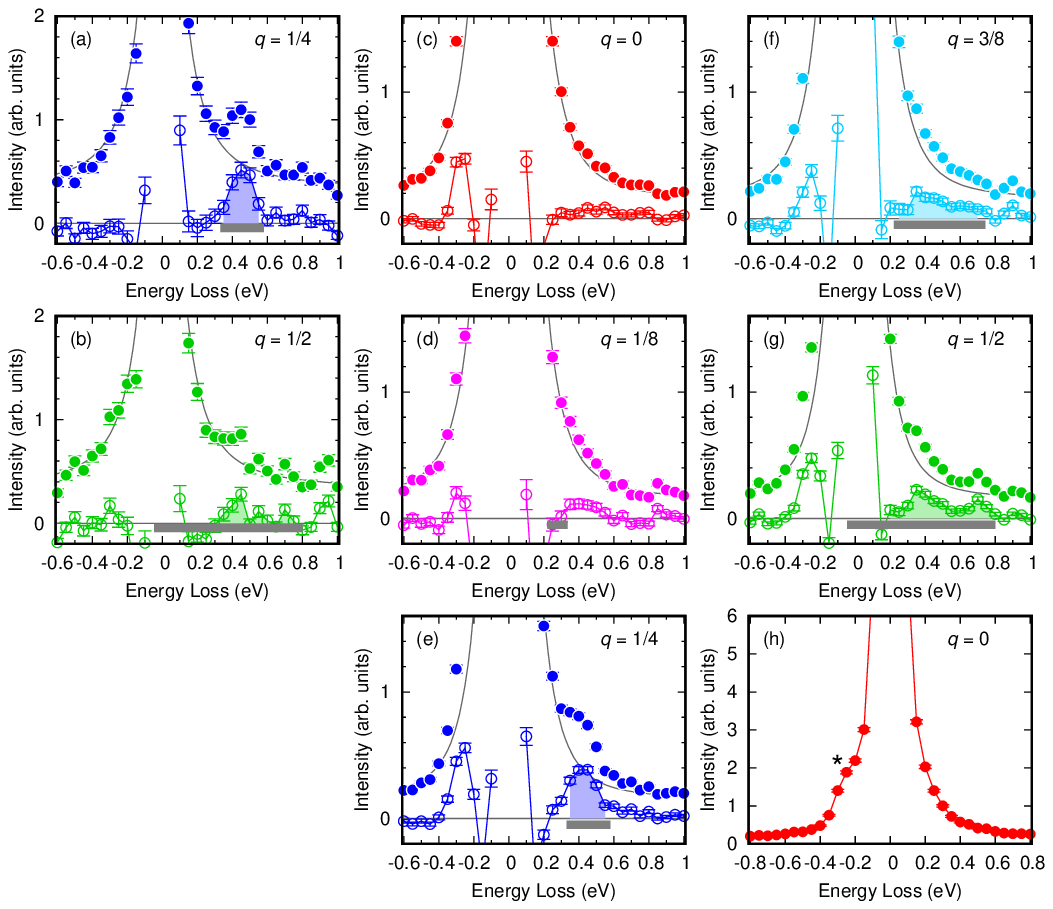}
\caption{Momentum dependence of the Cu $K$-edge RIXS spectra of Sr$_2$CuO$_3$.
The incident photon energy is $E_i=8982.5$ eV, and the absolute momentum transfer is $\bm{Q}=(0,q,3.56).$
The monochromator is Si(333) for (a)--(b) and Si(400) for (c)--(h).
The filled circles are the raw spectra.
In (a)--(g), the black solid lines represent the tail of the elastic scattering, and the open circles show the spectra after subtracting the tail.
The horizontal gray bars indicate the energy range of the two-spinon continuum broadened by the experimental energy resolution.
The spectrum in (h) is identical to that in (c), but plotted with different scales for energy loss and intensity to highlight the lineshape of tail of the elastic scattering.
A small bump, indicated by an asterisk, appears at –0.3 eV and causes a spurious peak in the subtracted spectra in (c)–(g).}
\label{fig:qdep}
\end{figure*}

A clear peak appears at 0.4 eV in configurations (a) and (b), which correspond to well-screened and poorly-screened resonances, respectively.
Since charge-transfer and orbital ($dd$) excitations are observed at higher energies ($>$1.5 eV) \cite{Hasan2002, Suga2005, Schlappa2012}, they are unlikely to be the origin of this peak.
In addition, the energy of 0.4 eV is too high to attribute the feature to a phonon mode.
Given the experimentally established value of $J$ = 0.24 eV for Sr$_2$CuO$_3$ \cite{Suzuura1996,Walters2009,Schlappa2012}, $E_{\rm u}$ and $E_{\rm l}$ are determined as 0.53 and 0.38 eV, respectively, at $q$ = 1/4.
Energy of the peak feature agrees very well with the two-spinon continuum.
Therefore, we ascribe the peak feature to the two-spinons excitations.
Because a single spin-flip excitation is forbidden in the Cu $K$-edge RIXS, the two-spinon excitations conserve the total spins ($\Delta S = 0$).
In the theory described later, the intensity of the poorly-screened resonance is larger than that of the well-screened resonance.
However, in the experiment, the observed intensities of the two resonances are not significantly different, probably due to large self-absorption at the poorly-screened resonance.
Since the quality of the data is better at the well-screened resonance, we choose this condition for the momentum scans in the next section.
In contrast to the spectra depicted in Figs.~\ref{fig:pol}(a) and (b), the peak feature is either absent or very weak in Fig.~\ref{fig:pol}(c), which demonstrates the polarization dependence of the two-spinon excitations.
The polarization dependence is discussed later.

\subsection{Momentum dependence}

Figure \ref{fig:qdep} shows the momentum dependence observed at the incident photon energy and polarization condition in Fig.~\ref{fig:pol}(a).
The momentum dependence was measured twice with different energy resolutions.
Figures \ref{fig:qdep}(a) and (b) show the spectra of the first measurement using the Si(333) monochromator at $q=1/4$ and $q=1/2$, respectively.
The spectrum in Fig.~\ref{fig:qdep}(a) is the same as that in Fig.~\ref{fig:pol}(a).
In contrast to the clear peak at $q=1/4$, the spectral weight observed below 1 eV is almost absent at $q=1/2$ after subtracting the tail of the elastic scattering.

Figures \ref{fig:qdep}(c)--(g) shows the second measurement of the momentum dependence using the Si(400) monochromator which gives a slightly poorer energy resolution than Si(333).
We note that a small bump exists in the anti-Stokes region, as indicated by the asterisk in Fig.~\ref{fig:qdep}(h), and it causes a spurious peak at $-0.3$ eV in the subtracted spectra (open circles) in Figs.~\ref{fig:qdep}(c)--(g).
In the Stokes region, the spectral weight at the energy range of the two-spinon continuum is observed clearly again at $q=1/2$.
The spectral weight in the two-spinon continuum is also observed at $q=3/8$.
In contrast, spectral intensity follows the tail of the elastic scattering at $q=0$ and $q=1/8$ and excitations are not observed.
Given that the excitations at low energy are buried in the huge elastic tail at $q=1/8$, the momentum dependence of the four momenta [Figs.~\ref{fig:qdep}(c)--(e)] is consistent with the two-spinon continuum induced by the $\Delta S = 0$ excitations described by the spin-exchange dynamical structure factor.

Theoretically, the spectral weight of the two-spinon continuum induced by the $\Delta S = 0$ excitation is exactly zero at $q=1/2$; however small spectral weight remains after subtracting the tail of the elastic scattering in Fig.~\ref{fig:qdep}(g).
The apparent discrepancy in the residual spectral weight between Figs.~\ref{fig:qdep}(b) and (g) may be attributed to insufficient statistical accuracy.
Additionally, the experimental momentum resolution, $\Delta q \simeq 0.09$, is low and insufficient to prevent contamination of the spectra, nominally measured at $q=1/2$, by the finite spectral weight at $q \ne 1/2$.
From these two measurements, it is reasonable to conclude that the momentum dependence of the Cu $K$-edge RIXS is consistent with 
the two-spinon continuum induced by $\Delta S=0$, namely, the excitations are prominent at $q=1/4$ and decrease in intensity toward both the zone center and the zone boundary.

\subsection{Theoretical calculation}

\begin{figure}[t]
\centering
\includegraphics[width=1.0\columnwidth]{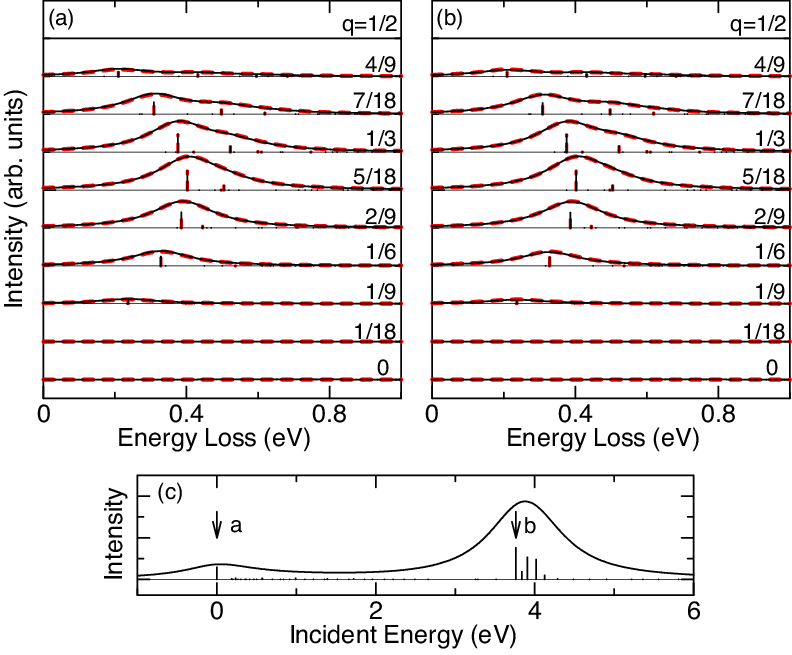}
\caption{(a) and (b) Calculated RIXS spectra below 1 eV (black solid lines) and the spin-exchange dynamical structure factor (red broken lines) in an 18-site extended Hubbard model.
The parameters used in the calculation are given in the text.
The $\delta$-function (vertical lines) is convoluted with the Lorentzian broadening of $0.2t$.
The intensity of each spectrum is normalized relative to the weight of the lowest-energy excitation at $q=5/18$, and the original intensity in (b) is approximately 5.5 times larger than that in (a).
(c) X-ray absorption spectrum of the model.
Lorentzian broadening of $t$ is applied.
The arrows indicate the incident photon energy for the calculation of RIXS in (a) and (b).}
\label{fig:theory}
\end{figure}

To support the experimental observation of the two-spinon continuum, we performed a theoretical calculation of the spin-exchange dynamical structure factor and the Cu $K$-edge RIXS spectra using the numerical exact diagonalization technique on an 18-site extended Hubbard chain.
This model serves as a minimal representation of one-dimensional cuprates \cite{Neudert1998,Tsutsui2000b,Kim2004} and is defined by the Hamiltonian,
\[
H = -t \sum_{i,\sigma} (c_{i,\sigma}^\dagger c_{i+1,\sigma} + \text{h.c.}) + U \sum_i n_{i\uparrow} n_{i\downarrow} + V \sum_{i} n_i n_{i+1},
\]
where $c_{i,\sigma}^\dagger$ creates electrons with spin $\sigma$ at site $i$ , $n_{i,\sigma}=c_{i,\sigma}^\dagger c_{i,\sigma}$, and $n_i=n_{i,\uparrow}+n_{i,\downarrow}$.
The parameters, $t$, $U$, and $V$ correspond to the nearest-neighbor hopping, the on-site Coulomb repulsion, and the nearest-neighbor Coulomb repulsion, respectively.
The last term is introduced to capture excitionic effects \cite{Neudert1998,Tsutsui2000b}, even though spin excitations remain qualitatively unchanged \cite{Kourtis2012}.
The spin-exchange dynamical structure factor is defined as \cite{Klauser2011a},
\[
S^\text{ex}(q,\omega) = \sum_f |\langle f | X_q | 0 \rangle|^2 \delta(\omega - E_f + E_0),
\]
where $X_q=\sum_j e^{iqj} (\mathbf{S}_{j-1}\cdot \mathbf{S}_j+\mathbf{S}_j\cdot \mathbf{S}_{j+1})$ is the spin-exchange operator, $\mathbf{S}_i$ is the spin operator at site $i$, and $|0\rangle$ and $|f\rangle$ are the ground and excited states with energy $E_0$ and $E_f$, respectively.
The Cu $K$-edge RIXS spectrum is expressed as a second-order dipole transition process between Cu $1s$ and $4p$ orbitals.
The intermediate state of the process involves the Coulomb interaction between $3d$ and $1s$-core holes, given by,
\(
H_{1s-3d}=-U_c \sum_{i,\sigma} n_i n_{i\sigma}^s,
\)
where $n_{i\sigma}^s$ is the number operator of $1s$-core hole.
By assuming that the $4p$ photo-electron enters into the bottom of the $4p$ band with momentum $k_0$, the RIXS spectrum is expressed as \cite{Tsutsui2000b},
\[
I(q,\omega) = \sum_f \left| \langle f | D_{q_f} ^\dagger G(\omega_i) D_{q_i} | 0 \rangle \right|^2 \delta(\omega - E_f + E_0),
\]
where $q=q_f-q_i$, $D_q =\sum_\sigma p_{k_0,\sigma}^\dagger s_{k_0-q,\sigma}^\dagger$, $s_{k,\sigma}^\dagger$ ($p_{k,\sigma}^\dagger$) is the creation operator of the $1s$ core hole ($4p$ electron) with momentum $k$ and spin $\sigma$, $G^{-1}(\omega_i)=H+H_{1s-3d}+\varepsilon_{1s-4p}-E_0-\omega_i-i\Gamma$, $\omega_i$ is the incident photon energy, $\Gamma$ is the inverse lifetime of the intermediate state, and  $\varepsilon_{1s-4p}$ is the energy difference between the $1s$ level and the bottom of the $4p$ band.

The parameters are set as $U/t=10$, $V/t=1.5$, and $U_c/t=15$ in accordance with Ref.\ \cite{Tsutsui2000b},
while the value $\Gamma/t=3$ was chosen to capture the intensity trend seen in the \(dp\) model \cite{Okada2006}.
The nearest-neighbor exchange interaction is given by $J=4t^2/(U-V)=0.24$ eV \textcolor{red}{\cite{Suzuura1996,Walters2009,Schlappa2012}}.
The numerical outcomes described below are not sensitive to the detailed choice of parameters.

The spin-exchange dynamical structure factor and the Cu $K$-edge RIXS spectra are compared in Figs.~\ref{fig:theory}(a) and (b), where the incident photon energies are set to the peaks indicated by the arrows and corresponding labels in the x-ray absorption spectrum shown in Fig.~\ref{fig:theory}(c).
Peaks a and b correspond to the well-screened and poorly screened states, respectively.
Our results are consistent with the spin-exchange dynamical structure factor of the Heisenberg model \cite{Forte2011a,Klauser2011a}, as well as  RIXS spectra of the magnetic excitations in the Hubbard and Heisenberg models \cite{Kourtis2012,Igarashi2012a}.
We note here that our direct comparison demonstrates excellent agreement between the spin-exchange dynamical structure factor and Cu $K$-edge RIXS spectra when both are calculated using the same model, confirming that the experimentally observed spectral weight below 1 eV originates from the two-spinon continuum.

In the Cu $K$-edge RIXS, the interaction between the $1s$ core hole and the valence electrons in the intermediate state plays a crucial role in inducing electronic excitations.
Although the excitation process is complicated in the Cu $K$-edge RIXS, a similarity between the RIXS spectra and dynamical structure factor has been argued.
In the two-dimensional undoped Hubbard model, the spectral weight of the Mott-gap excitations in the Cu $K$-edge RIXS appears in the same energy range as the dynamical charge structure factor, but their spectral lineshapes differ \cite{Jia2012}.
In the doped case, the Cu $K$-edge RXIS spectra are found to be similar to the dynamical charge structure factor, when the core hole is created at a doped site by tuning the incident photon energy \cite{Ishii2005b,Jia2012}.
Whereas this similarity to the charge structure factor is only qualitative, the present calculations show more pronounced similarity to the spin exchange dynamical structure factor; the calculated RIXS spectra are almost identical to the spin exchange dynamical structure factor.
Moreover, this similarity holds for both the incident photon energies corresponding to the well-screened (a) and poorly-screened (b) states.

\section{Discussion}

Our study demonstrates that the Cu $K$-edge RIXS offers a new opportunity to investigate the fractionalization of elementary excitations and novel quantum state in one-dimensional systems.
Interestingly, apart from the differences in intensity distribution, Cu $K$-edge RIXS detects spin excitations in the same energy-momentum space as the Cu $L$-edge RIXS in the spin-flip channel and INS in the one-dimensional spin system.
This stands in stark contrast to two-dimensional spin systems, where single-magnon excitations dominate the low energy region in the Cu $L$-edge RIXS and INS, while such excitations are forbidden in the Cu $K$-edge RIXS.

The O $K$-edge RIXS can serve as an alternative technique to measure the spin-exchange dynamical structure factor.
As long as the spin-orbit interaction of the valence electrons can be neglected, spin-flip excitations are forbidden in the O $K$-edge RIXS \cite{Lu2018}, which is indeed the case for copper oxides.
The two-spinon excitations in Sr$_2$CuO$_3$ have been observed using the O $K$-edge RIXS \cite{Schlappa2018}; however, the observation is limited to regions near $q=0$, and a momentum constraint prevents access to the vicinity of $q=1/2$, where the intensity of the two-spinon continuum differs between the $\Delta S=0$ and $\Delta S=1$ channels.
Additionally, a theoretical study based on the half-filled $t-J$ model \cite{Kumar2018} shows that the intensity in the two-spinon continuum does not vanish at $q=1/2$ due to electron hopping from a core-hole site under the weak core-hole potential of the oxygen $1s$ orbital.
This means that RIXS spectra at the O $K$-edge deviate from the spin exchange dynamical structure factor.
In these senses, Cu $K$-edge RIXS is more suitable for measuring the spin-exchange dynamical structure factor.
If the spin-conserving ($\Delta S=0$) and spin-flip ($\Delta S=1$) channels can be separated by analyzing the polarization of the scattered photons\textcolor{red}{.}
The use of polarization-resolved RIXS has been gradually expanding recently \cite{Betto2021,Martinelli2022},
it has not yet been applied to one-dimensional systems.

At the zone center in Fig.~\ref{fig:qdep}(c), no inelastic signal is detected at the Cu $K$ edge, whereas four-spinon excitations are observed around 0.7 eV at the O $K$ edge \cite{Schlappa2018} and at the Cu $L_3$ edge \cite{Kumar2022}, although they are less pronounced at the latter edge.
In principle, four-spinon excitations are allowed at the Cu $K$-edge, but their occurrence requires a long core-hole lifetime, as discussed in Ref.~\cite{Schlappa2018}.
Since the lifetime of the Cu $K$-level is an order of magnitude shorter than that of the O $K$-level \cite{Campbell2001}, four-spinon excitations are not easily observable in Cu $K$-edge RIXS.
Excellent agreement between the Cu $K$-edge RIXS spectra and the spin-exchange dynamical structure factor obtained in our calculations supports lack of four-spinon excitations in the experimental spectrum.

Finally, we discuss the weak or absent signal of the two-spinon excitations shown in the spectrum in Fig.~\ref{fig:pol}(c).
In the transition-metal $K$-edge RIXS, a phenomenological analysis of the symmetry between the polarization and the $d$ orbital yields a necessary condition for observing the $dd$ excitations \cite{Ishii2011a}: at least one common symmetry should be shared in the reduced product-representations $\Gamma_i \times \Gamma_f$ and $P_i \times P_f$.
Here, $\Gamma_{i(f)}$ and $P_{i(f)}$ are the irreducible representations of the initial (final) electronic orbital and those of the incident (scattered) photon polarization, respectively.
We approximate the local symmetry of the Cu atom as $D_{4h}$, rather than the exact site symmetry of $D_{2h}$, because the Cu–O distances along the $a$ and $b$ axes are nearly equal \cite{Ami1995}.
Using a coordinate system in which the $x$, $y$, and $z$ axes are parallel to the $a$, $b$, and $c$ axes, respectively, the symmetry of $P_i \times P_f$ should include $A_{1g}$ ($\Gamma_{x^2-y^2} \times \Gamma_{x^2-y^2}$) because the two-spinon excitation occurs within the $d_{x^2-y^2}$ orbital.
In the configuration of Figs.~\ref{fig:pol} (a) and (b), the component of ${\boldsymbol \epsilon}_\mathrm{i} \parallel {\bm z}$ is relevant for the resonance and scattered photon polarization ${\boldsymbol \epsilon}_\mathrm{f}$ is parallel to either ${\bm y}-{\bm z}$ or ${\bm x}$.
Then, $P_i \times P_f$ is $A_{1g}+E_g$, and the two-spinon excitations are allowed.
The observation of excitations at both the well-screened [Fig.~\ref{fig:pol}(a)] and poorly-screened [Fig.~\ref{fig:pol}(b)] intermediate states is qualitatively consistent with the theoretical calculations shown in Figs.~\ref{fig:theory}(a) and (b).
On the other hand, in the configuration of Figs.~\ref{fig:pol} (c), ${\boldsymbol \epsilon}_\mathrm{i}$ is parallel to  ${\bm x}+{\bm y}$ and ${\boldsymbol \epsilon}_\mathrm{f}$ is parallel to either ${\bm x}-{\bm y}$ or ${\bm z}$.
Since $P_i \times P_f$ is $B_{1g}+A_{2g}+E_g$, the two-spinon excitations are forbidden, which is consistent with our observation.

\section{Conclusion}
In this study, we observed spin excitations in the $S=1/2$ antiferromagnetic chain system Sr$_2$CuO$_3$ using the Cu $K$-edge RIXS.
The spin excitations appear in the two-spinon continuum, and the spectral weight of the spin excitations is prominent near the midpoint between the center and the boundary of the Brillouin zone.
The observed momentum dependence aligns with the spin-exchange dynamical structure factor, demonstrating the potential of the $K$-edge RIXS as a tool for observing a novel spin dynamics in one-dimensional systems.
The spin excitations are observed under the resonant condition of ${\boldsymbol \epsilon}_\mathrm{i} \parallel {\bm c}$ but are almost absent when ${\boldsymbol \epsilon}_\mathrm{i} \parallel {\bm a}, {\bm b}$.
This polarization dependence can be interpreted through a phenomenological analysis of the symmetry relationship between the polarization and the $d$ orbital.

\begin{acknowledgments}
This work was financially supported by JSPS KAKENHI Grants No.~16H04004 and No.~24K00560.
The synchrotron radiation experiments at SPring-8 were carried out at the BL11XU with the approval of the Japan Synchrotron Radiation Research Institute (JASRI) (Proposals No.~2018B3555, No.~2019A3555, No.~2020A3555, and No.~2021A3555).
The numerical calculation was partially carried out on the supercomputing facilities in National Institutes for Quantum Science and Technology (QST).
\end{acknowledgments}

\section*{Data Availability}
The data that support the findings of this article are openly available \cite{opendata}.

\bibliography{sr2cuo3-paper}

\end{document}